\documentstyle[12pt,prd,aps,epsfig]{revtex}
\tighten

\title{Top quark associated production of the neutral top-pion at high energy
$e^{+}e^{-}$ colliders}

\author{Chongxing Yue$^{(a,b)}$, Yong Jia$^{b}$, Yanming Zhang$^{b}$, Hong Li$^{b}$\\
{\small a: CCAST(World Laboratory) P.O. BOX 8730. B.J.
100080P.R.China}\\
{\small b: College of Physics and Information
Engineering,}\\
\small{Henan Normal University, Xinxiang  453002.
P.R.China}
\thanks{This work is supported by the National Natural Science
Foundation of China(I9905004), the Excellent Youth Foundation of
Henan Scientific Committee(9911), and Foundation of Henan
Educational Committee.}
\thanks{E-mail:cxyue@public.xxptt.ha.cn} }
\date{\today}
\begin{document}
\maketitle
\begin{abstract}
\hspace{5mm} In the context of topcolor-assisted technicolor (TC2)
models, we calculate the associated production of the neutral
top-pion $\pi_{t}^{0}$ with a pair of top quarks via the process
$e^{+}e^{-}\longrightarrow t\bar{t}\pi_{t}^{0}$. We find that the
production  cross section is larger than that of the process  $
e^{+}e^{-}\longrightarrow t\bar{t}H$ both in the standard model
(SM) and in the minimal supersymmetric SM. With reasonable values
of the parameters in TC2 models, the cross section can reach
$20fb$. The neutral top-pion $\pi_{t}^{0}$ may be direct observed
via this process.

\end {abstract}

\vspace{1.0cm} \noindent
 {\bf PACS number(s)}: 12.60.Fr, 14.80.Cp, 12.60.Rc

\newpage
The mechanism of electroweak symmetry breaking (EWSB) remains the
most prominent mystery in elementary particle physics despite the
success of the standard model (SM) tested by the high energy
experimental data. The present and next generation of colliders
will help explaining the nature of the EWSB and the origin of
fermion masses. The LHC is expected to directly probe possible new
physics (NP) beyond the SM up to a scale of a few TeV, while the
high energy linear $e^{+}e^{-}$ collider (LC) is required to
complement the probe of the new particles with detailed
measurements. Further more, some kinds of NP predict the existence
of new particles that would be manifested as rather spectacular
resonance in the LC experiments, if the achievable centre-of-mass
energy is sufficient. Even their masses exceed the centre-of-mass
energy, it also retains an indirect sensitivity through a
precision study of the virtual corrections to electroweak
observable.  A LC represents an ideal laboratory for studying this
kind of NP\cite{y1}.

Until a Higgs boson with large coupling to gauge boson pair is
discovered, the possibility of strong EWSB must be
entertained\cite{y2}. The most commonly studied class of theories
is technicolor (TC)\cite{y3}, which dynamical break the
electroweak symmetry. Although TC models have many theoretical
problems as well as conflicts with data and broad classes of these
models have been ruled out, there are still viable models worthy
of investigation in light of the capabilities of the current
generation of experiments.

It is widely believed that the top quark will be a sensitive probe
into physics beyond the SM.The properties of the top quark could
reveal information about flavor physics, EWSB, as well as
NP\cite{y4}. Given the large value of the top quark mass and the
sizable splitting between the masses of the top and bottom quarks,
it is natural to wonder whether $m_{t}$ has a different origin
from the masses of the other quarks and leptons. There may be a
common origin for EWSB and top quark mass generation. Much
theoretical work has been carried out in connection to the top
quark and EWSB. The TC2 models\cite{y5}, the top see-saw
models\cite{y6}, and the flavor universal TC2 models\cite{y7} are
three of such examples. Such type of models generally predict a
number of scalars with large Yukawa coupling to the third
generation. For example, TC2 models predict the existence of the
top-pions ($\pi_{t}^{\pm}$, $\pi_{t}^{0}$). These new particles
are most directly related to the EWSB. Thus, studying the possible
signatures of these new particles at present and future high
energy colliders would provide crucial information for the EWSB
and can be used to test TC2 theory.

The production of a Higgs boson predicted by the SM or the minimal
supersymmetric SM (MSSM) in association with a pair of top-antitop
quarks is of extreme interest for two reasons\cite{y8}. First, the
$t\bar{t}H$ production mode can be important for discover of a
Higgs boson around $120-130GeV$ at the LHC, and even at the Run II
of the Tevatron with high enough luminosity\cite{y9}. Although it
has a small event rate($1\sim 5fb$) for a SM-like Higgs boson,and
even lower for an MSSM Higgs boson, the signature is quite
spectacular. Second, this production mode offers a direct handle
on the Yukawa coupling of the top quark, supposedly the most
relevant one to understand the nature of fermion masses. Both the
LHC and LC can use the second feature to try a precision
measurement of the $t\bar{t}H$ coupling.

The signals of the $t\bar{t}H$ production mode has been studied
quite extensively in the LC\cite{y10}, Tevatron and LHC\cite{y9}
in the context of the SM and MSSM. Recently, Ref\cite{y11} has
investigated the associated production of a neutral scalar
predicted by TC2 models with a pair of top quarks at hadron
colliders. They find that the neutral scalar may be observed at
the LHC via the process $p\bar{p}\rightarrow t\bar{t}\bar{\phi}$.
In this paper, we will calculate the production cross section of
the process $e^{+}e^{-}\rightarrow t\bar{t}\pi_{t}^{0}$ and the
discuss the possibility of the detecting $\pi_{t}^{0}$ via this
process in the future LC experiments. We find the production cross
section is larger than that of the process $e^{+}e^{-}\rightarrow
t\bar{t}H$. In most of the parameter space of the TC2 models, the
production cross section is larger than $5fb$ which may be
detected in the future LC experiments.

TC2 models generate the large top quark mass through the formation
of a dynamical $t\bar{t}$ condensation and provide possible
dynamical mechanism for breaking electroweak symmetry. In TC2
models, the EWSB is driven mainly by TC interaction, ETC
interactions give contribution to all ordinary quark and lepton
masses including a very small portion of the top quark mass,
namely $m_{t}^{\prime}=\epsilon m_{t}$ with $\epsilon\ll 1$. The
topcolor interactions also make small contributions to the EWSB
and give rise to the main part of the top quark mass
$m_{t}-m_{t}^{\prime}=(1-\epsilon)m_{t}$ similar to the
constituent masses of the light quark in QCD. This means that the
associated top-pions $\pi_{t}^{0}$, $\pi_{t}^{\pm}$ are not the
longitudinal bosons W and Z, but separately, physically observable
objects.

To maintain eletroweak symmetry between top and bottom quarks and
get not generate $m_{b}\approx m_{t}$, the topcolor gauge group is
usually take to be a strongly coupled $SU(3)\bigotimes U(1)$. At
the $\Lambda\sim 1TeV$, the dynamics of a general TC2 model
involves the following structure\cite{y5,y12}:
\begin{equation}
SU(3)_{1}\bigotimes SU(3)_{2}\bigotimes U(1)_{Y_{1}}\bigotimes
U(1)_{Y_{2}}\bigotimes SU(2)_{L} \longrightarrow
SU(3)_{QCD}\bigotimes U(1)_{EM},
\end{equation}
where $SU(3)_{1}\bigotimes U(1)_{Y_{1}} (SU(3)_{2}\bigotimes
U(1)_{Y_{2}})$ generally preferentially to the third (first and
second) generations. The $U(1)_{Y_{2}}$ are just strongly rescaled
versions of electroweak $U(1)_{Y}$. The above breaking scenario
gives rise to the topcolor gauge bosons including the color-octet
coloron $B_{\mu}^{A}$ and color-singlet extra $U(1)$ gauge bosons
$Z^{\prime}$. The relevant couplings of the new gauge boson
$Z^{\prime}$ to ordinary fermions can be written as:
\begin{equation}
\frac{1}{2}g_{1}[\frac{1}{3} \cot\theta^{\prime}
Z_{\mu}^{\prime}(\bar{t}_{L}\gamma^{\mu}t_{L}+2\bar{t}_{R}\gamma^{\mu}t_{R})
+\tan\theta^{\prime}Z_{\mu}^{\prime}(\bar{e}_{L}\gamma^{\mu}e_{L}+2\bar{e}_{R}
\gamma^{\mu}e_{R})],
\end{equation}
where $g_{1}$ is the $U(1)_{Y_{1}}$ coupling constant at the scale
$\Lambda_{TC}$ and $\theta^{\prime}$ is the mixing angle with
$\tan\theta^{\prime}=g_{1}/(2\sqrt{\pi k_{1}})$. To obtain proper
vacuum tilting (the topcolor interactions only condense the top
quark but not the bottom quark), the coupling constant $k_{1}$
should satisfy certain constraint, ie. $k_{1}\leq 1$\cite{y7,y12}.
In the following estimation, we will take $k_{1}=1$. The choice
$k_{1}=1$ corresponds to $\tan^{2}\theta^{\prime}=0.01$\cite{y12}.

For TC2 models, the underlying interactions, topcolor
interactions, are non-universal and therefore do not posses a GIM
mechanism. The non-universal gauge interactions can result in the
flavor changing (FC) coupling vertices when one writes the
interactions in the quark mass eigen-basis. Thus, the top-pions
have large Yukawa couplings to the third family quarks and can
induce the new FC scalar couplings. For the neutral top-pion
$\pi_{t}^{0}$, the relevant coupling can be written
as\cite{y5,y13}:
\begin{equation}
\frac{m_{t}}{\sqrt{2}F_{t}}\frac{\sqrt{\nu_{w}^{2}-F_{t}^{2}}}{\nu_{w}}
[K_{UR}^{tt}K_{UL}^{tt\ast}\bar{t}_{L}t_{R}\pi_{t}^{0}
+K_{UR}^{tc}K_{UL}^{tt\ast}\bar{t}c_{R}\pi_{t}^{0}+h.c.],
\end{equation}
where $F_{t}$ is the top-pion decay constant and
$\nu_{w}=\nu/\sqrt{2}=174GeV$. It has been shown that the values
of these coupling parameters can be taken as:
\begin{equation}
K_{UR}^{tt}=K_{DL}^{bb}=1,\hspace{5mm}K_{UR}^{tt}=1-\epsilon,
\hspace{5mm}K_{UR}^{tc}\leq
\sqrt{2\epsilon-\epsilon^{2}}.
\end{equation}
From above discussion, we can see that the $t\bar{t}\pi_{t}^{0}$
production channels can be represent by the processes:
\begin{equation}
e^{+}e^{-}\rightarrow \gamma^{\ast}, Z^{\ast}, Z^{\prime
\ast}\rightarrow t\bar{t}\pi_{t}^{0}
\end{equation}
Using Eq.(2)-(4) and other relevant Feynnman rules, we can give
the invariant scattering amplitude of the process
$e^{+}e^{-}\rightarrow t\bar{t}\pi_{t}^{0}$.
\begin{equation}
M=M_{\gamma}+M_{Z}+M_{Z^{\prime}},
\end{equation}
with
\begin{eqnarray}
M_{\gamma} & = & \bar{v}(p_{-})
(-i)e\gamma_{\mu}u(p_{+})\frac{-ig^{\mu\nu}}{s} \nonumber
\\
           &   & \times\frac{i}{\rlap/p_{t}+\rlap/p_{\pi_{t}^{0}}-m_{t}}
           \bar{u}(p_{t})i\beta\gamma_{5}ie\frac{2}{3}
\gamma_{\nu}v(p_{\bar{t}}) , \\
 M_{Z} & = &
\bar{v}(p_{-})i\gamma_{\mu}(v_{e}+a_{e}\gamma_{5})
u(p_{+})\frac{-ig^{\mu\nu}}{s-m_{Z}^{2}+im_{Z}\Gamma_{Z}}
\nonumber
\\
           &   & \times\frac{i}
{\rlap/p_{t}+\rlap/p_{\pi_{t}^{0}}-m_{t}}\bar{u}
(p_{t})i\beta\gamma_{5}i\gamma_{\nu}
(v_{t}+a_{t}\gamma_{5})v(p_{\bar{t}}) , \\
 M_{Z^{\prime}} & = &
\bar{v}(p_{-})i\gamma_{\mu}(v_{e}^{\prime}+a_{e}^{\prime}
\gamma_{5})u(p_{+})\frac{-ig^{\mu\nu}}{s-m_{Z^{\prime}}^{2}+im_{Z^{\prime}}\Gamma_{Z^{\prime}}}
\nonumber
\\
                &   & \times\frac{i}
{\rlap/p_{t}+\rlap/p_{\pi_{t}^{0}}-m_{t}}\bar{u}(p_{t})
i\beta\gamma_{5}i\gamma_{\nu}
(v_{t}^{\prime}+a_{t}^{\prime}\gamma_{5})v(p_{\bar{t}}) ,
\end{eqnarray}
where
\begin{eqnarray}
\beta & = &
\frac{m_{t}}{\sqrt{2}F_{t}}\frac{\sqrt{\nu_{w}^{2}-F_{t}^{2}}}{\nu_{w}}K_{UR}^{tt}
K_{UL}^{tt\ast}
,\hspace{5mm}
 v_{e}  =
 \frac{e}{4\sin\theta\cos\theta}(4\sin^{2}\theta-1),\\
 a_{e} & = &
 \frac{e}{4\sin\theta\cos\theta},\hspace{30mm}
 v_{t}  =
 \frac{e}{4\sin\theta\cos\theta}(1-\frac{8}{3}\sin^{2}\theta),\\
 a_{t} & = &
 -\frac{e}{4\sin\theta\cos\theta},\hspace{27mm}
 v_{e}^{\prime}  =
 \frac{3}{4}\sqrt{4\pi k_{1}}\tan^{2}\theta^{\prime},\\
 a_{e}^{\prime} & = &
 \frac{-1}{4}\sqrt{4\pi k_{1}}\tan^{2}\theta^{\prime},\hspace{10mm}
 v_{t}^{\prime}  =
 \frac{5}{12}\sqrt{4\pi k_{1}},\hspace{10mm}
 a_{e}^{\prime}  =
 \frac{-1}{4}\sqrt{4\pi k_{1}}.
\end{eqnarray}

Where $\sqrt{s}$ is the centre-of-mass energy of the LC
experiments and $M_{Z^{\prime}}$ is the mass of the topcolor gauge
boson $Z^{\prime}$. The decay width $\Gamma_{Z^{\prime}}$ is
dominated by $t\bar{t}$, $b\bar{b}$ and we have\cite{y14}:
\begin{equation}
\Gamma_{Z^{\prime}}\approx
\frac{g_{1}^{2}\cot^{2}\theta^{\prime}}{12\pi}M_{Z^{\prime}}=\frac{1}{3}M_{Z^{\prime}},
\end{equation}
which corresponds to $\tan^{2}\theta^{\prime}=0.01$.

To obtain numerical results we take the SM parameters as
$\sin^{2}\theta=0.2315$, $\alpha=\frac{1}{128.8}$,
$m_{Z}=91.2GeV$, $\Gamma_{Z}=2.495GeV$ and
$m_{t}=175GeV$\cite{y15}. For TC2 models, the parameters
$m_{\pi_{t}}$, $\epsilon$ and $M_{Z^{\prime}}$ are the free
parameters. To see how those parameters affect our numerical
results, we take $200GeV\leq m_{\pi_{t}}\leq450GeV$,
$0.01\leq\epsilon\leq 0.1$, and $2TeV\leq M_{Z^{\prime}}\leq
6TeV$.

The $t\bar{t}\pi_{t}^{0}$ production cross section $\sigma$ is
plotted in Fig.1 as a function of $m_{\pi_{t}}$ for the
centre-of-mass energy $\sqrt{s}=1.5TeV$, $M_{Z^{\prime}}=2.5TeV$
and three values of the parameter $\epsilon$. From Fig.1 we can
see that $\sigma$ is not sensitive to the value of parameter
$\epsilon$ and decreases with $m_{\pi_{t}}$ increasing. Thus, in
the following numerical estimation, we will take $\epsilon=0.05$.
For $\epsilon=0.05$, $\sigma$ varies between $10.7fb$ and $48fb$
for the neutral top-pion mass $m_{\pi_{t}}$ in the range of
$200-400GeV$.

To see the effect of the extra $U(1)_{Y}$ gauge bosons
$Z^{\prime}$ mass $M_{Z^{\prime}}$ on the cross section $\sigma$,
we plot the $\sigma$ as function of $M_{Z^{\prime}}$ for
$\sqrt{s}=1.5TeV$ in Fig.2, in which the full, broken and
dotted-dashed curves stand for $m_{\pi_{t}}=250GeV$, $350GeV$, and
$450GeV$, respectively. One can see from Fig.2 that $\sigma$
decreases with $m_{Z^{\prime}}$ increasing. In most of the
parameter space, the production cross section $\sigma$ is larger
than $5.8fb$.

Recently, Ref.[16] has shown that $B\bar{B}$ mixing provides
strong lower bounds on the masses of $B_{\mu}^{A}$ and
$Z^{\prime}$ bosons, i.e. there must be larger than $4TeV$. In
Fig.3 we take $M_{Z^{\prime}}=5TeV$ and plot $\sigma$ as a
function of $\sqrt{s}$ for three values of the top-pion mass. From
Fig.3 we can see that, when $\sqrt{s}$ is larger than $2TeV$, the
cross section $\sigma$ is not sensitive to the value of the
centre-of-mass energy $\sqrt{s}$. For $\sqrt{s}=2TeV$ and
$\epsilon=0.05$, $\sigma$ increase from $6.8fb$ to $21.8fb$ as
$m_{\pi_{t}}$ decreasing from $450GeV$ to $250GeV$.

The process $e^{+}e^{-}\rightarrow t\bar{t}H$ has been extensively
studied and calculated both in the SM and in the MSSM at
$O(\alpha_{s})$\cite{y10}. The cross section turns out to be
highly sensitive to the top Yukawa coupling $g_{t\bar{t}H}$ over
most of the parameter space. Although the cross section is very
smaller than $2fb$ for $m_{H}=120-130GeV$, the signature for
$t\bar{t}H$ production mode is spectacular. It has been shown that
the top Yukawa coupling $g_{t\bar{t}H}$ can be measured with a
precision of the order of $7-15\%$ at $\sqrt{s}=1TeV$ for
$M_{H}=120-130GeV$, assuming a b-tagging efficiency between 0.6
and 1\cite{y17} and with a precision of $5.5\%$ at
$\sqrt{s}=800GeV$, when optimal b-tagging efficiency is
assumed\cite{y18}. The cross section of the process
$e^+e^-\rightarrow t\bar{t}\pi_t^0$ is larger than $5fb$, even for
$M_{Z^{\prime}}=5$TeV and $\epsilon=0.01$, which is significantly
larger than that of the process $e^+e^-\rightarrow t\bar{t}H$.
Thus, the top Yukawa coupling $g_{t\bar{t}\pi_t^0}$ can be
precisely measured in the future LC experiments. The signatures
for $t\bar{t}\pi_t^0$ production mode may be detected.

Since the negative top-pion corrections to the $Z\rightarrow
b\bar{b}$ branching $R_{b}$ become smaller when the top-pion is
heavier, the LEP/SLD data of $R_{b}$ give lower bound on the
top-pion mass\cite{y19}. Ref.[20] has shown that the top-pion mass
is allowed to be in the range of a few hundred GeV depending on
the values of the parameters in the TC2 models. So we have take
the mass of the top-pion to vary in the range of $200GeV-450GeV$.
For $200GeV\leq m_{\pi_{t}}\leq 350GeV$, the dominate decay
channel is $\pi_{t}^{0}\rightarrow \bar{t}c$. If we take
$K_{UR}^{tc}=0.05$, $m_{\pi_{t}}=300GeV$, we have that the
branching ratio $B_{r}(\pi_{t}^{0}\rightarrow \bar{t}c)$ is
approximately equal to $60\%$ and $B_{r}(\pi_{t}^{0}\rightarrow
gg)$ is only equal to $16\%$\cite{y21}. This is different from
that of Ref.[11], which has assumed that the top-pion can not
induce the FC scalar coupling. Thus, the produce mode
$t\bar{t}\pi_{t}^{0}$ should be detected via the dominate decay
channel $\pi_{t}^{0}\rightarrow \bar{t}c$, which has been
extensively studied in Ref.[13]. For $m_{\pi_{t}}> 350GeV$, the
branching ratio of the decay channel $\pi_{t}^{0}\rightarrow
t\bar{t}$ is close to $100\%$, all other decay channels may be
ignored. In this case, the associated production of the
$\pi_{t}^{0}$ with a pair of top quarks induces an four top quark
final state, which may be experimentally observable\cite{y22}.

From triviality, we see that the SM can only be an effective
theory valid below some high energy scale $\Lambda$, the strong
EWSB theories might be needed\cite{y23}. The strong top dynamics
models, such as TC2 models, are the modern dynamical models of
EWSB. Such type of models generally predict a number of scalars
with large top Yukawa couplings. Direct observation of these new
particles via their large top Yukawa coupling would be
confirmation that the EWSB sector realized in nature is not the SM
or part of the MSSM. In this paper, we investigate the top quark
associated production with the neutral top-pion $\pi_t^0$ and
calculated the cross section of the $t\bar{t}\pi_t^0$ production
mode via the process $e^+e^-\rightarrow t\bar{t}\pi_t^0$ in the
context of TC2 models. Our results show that the cross section is
significantly larger than that of the process $e^+e^-\rightarrow
t\bar{t}H$ and is larger $5fb$ in most of the parameter space.
With reasonable values of the parameters in TC2 models, the cross
section can reach $20fb$. Thus, the neutral top-pion $\pi_t^0$ may
be direct observed via the process $e^+e^-\rightarrow
t\bar{t}\pi_t^0$.

\newpage
\begin{center}
{\bf Figure captions}
\end{center}
\begin{description}
\item[Fig.1:]The cross section $\sigma$ of the process
              $e^{+}e^{-}\longrightarrow t\bar{t}\pi_{t}$
              as a function of the mass $m_{\pi_{t}}$
              for $\sqrt{s}=1.5TeV$, $M_{Z^{\prime}}=2.5TeV$ and $\epsilon=0.03$ (full curve),
              $0.05$ (broken curve), and $0.08$ (dotted-dashed curve).
\item[Fig.2:]The cross section $\sigma$ versus the mass of $M_{Z^{\prime}}$
             for $\epsilon=0.05$, $\sqrt{s}=1.5TeV$ and $m_{\pi_{t}}=250GeV$ (full curve),
             $m_{\pi_{t}}=350GeV$ (broken curve)
             and $m_{\pi_{t}}=450GeV$ (dotted-dashed curve).
\item[Fig.3:]The cross section $\sigma$ versus the center-of mass $\sqrt{s}$
             for $\epsilon=0.05$, $M_{Z^{\prime}}=5TeV$ and $m_{\pi_{t}}=250GeV$ (full curve),
             $m_{\pi_{t}}=350GeV$ (broken curve)
             and $m_{\pi_{t}}=450GeV$ (dotted-dashed curve).

\end{description}

\newpage
\begin{figure}[hb]
\begin{center}
\epsfig{file=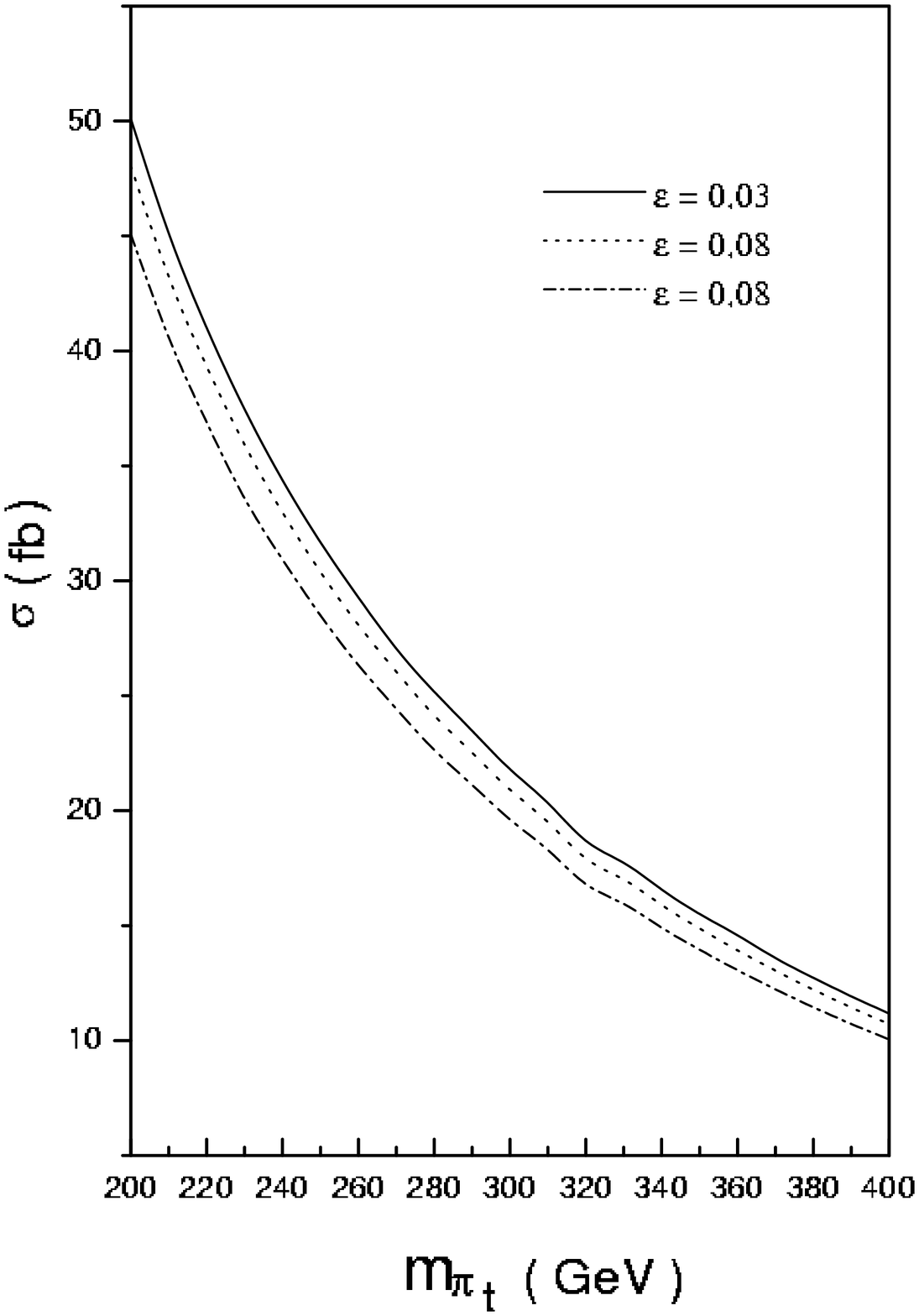,width=370pt,height=275pt}
\end{center}
\end{figure}
\begin{center}Fig.1
\end{center}

\newpage
\begin{figure}[hb]
\begin{center} \epsfig{file=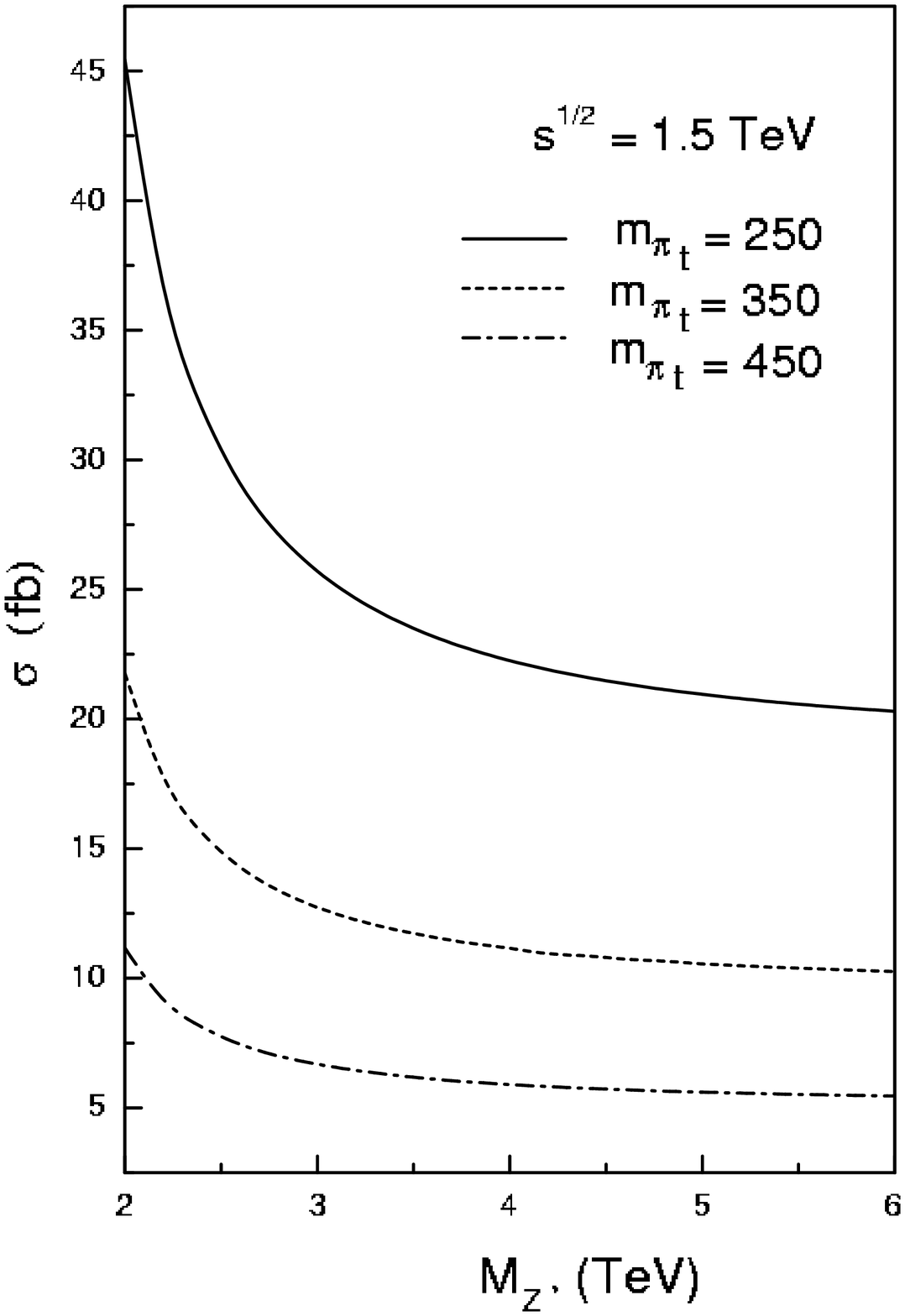,width=370pt,height=275pt}
\end{center}
\end{figure}
\begin{center}Fig.2
\end{center}

\newpage
\begin{figure}[hb]
\begin{center}
\epsfig{file=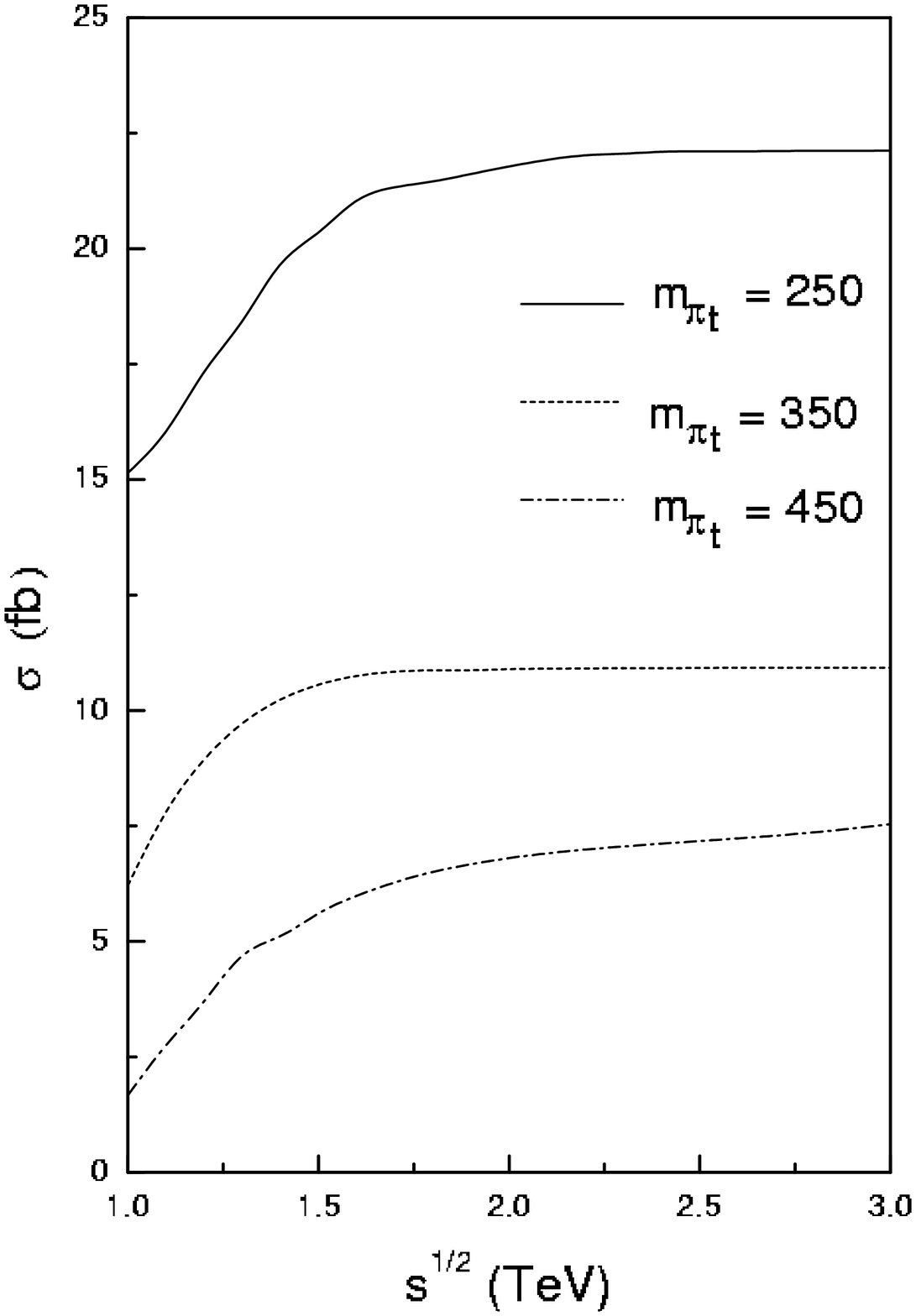,width=370pt,height=275pt}
\end{center}
\end{figure}
\begin{center}Fig.3
\end{center}
\end{document}